\documentclass[paper=a4, fontsize=12pt]{scrartcl}
\usepackage[margin=0.4in]{geometry}
\usepackage[T1]{fontenc}
\usepackage{fourier}
\usepackage{float}
\usepackage[english]{babel}	
\usepackage{caption}
\usepackage{relsize}
\usepackage{titlesec}
\usepackage[overload]{textcase}	
\usepackage{natbib}
\usepackage{xcolor}
\usepackage{braket}
\usepackage{moresize}
\usepackage{amsmath, amssymb, graphicx}
\usepackage{subcaption}
\usepackage{setspace}
\usepackage{natbib}
\usepackage[utf8]{inputenc}
\usepackage[protrusion=true,expansion=true]{microtype}	
\usepackage{amsmath,amsfonts,amsthm} 
\usepackage{graphicx}
\graphicspath{ {./images/} }	
\usepackage{url}

\usepackage{sectsty}
\allsectionsfont{\centering \normalfont\scshape}

\usepackage{fancyhdr}
\numberwithin{equation}{section}		
\numberwithin{figure}{section}			
\numberwithin{table}{section}				

\newcommand{\horrule}[1]{\rule{\linewidth}{#1}} 	

\title{
	\usefont{OT1}{bch}{b}{n}
	\horrule{0.5pt} \\[0.4cm]
	\huge First principles study of thermoelectric properties of $\text{Nb}_2\text{Co}_2\text{InSb}$ and $\text{Nb}_2\text{Co}_2\text{GaSb}$ double half-Heuslers \\
	\horrule{2pt} \\[0.5cm]
}
\author{
	\normalfont 								\normalsize
	Rajeev Ranjan\\[-3pt]		\normalsize
	\today
}
\date{}

\begin{document}

	\maketitle

	  \section*{ABSTRACT}
	  Valence electron count (VEC) 18 half-Heusler (hH) alloys are promising candidates
	  for thermoelectric applications at high temperatures due to their excellent Seebeck
	  coefficient, mechanical robustness, and thermal stability. However, their high lattice
	  thermal conductivity$(k_{L})$ poses a significant challenge for thermoelectric efficiency.Introduction of mass disorder at lattice sites can be an effective strategy to reduce the $k_{L}$ by scattering of heat carrying phonons. NbCoSn is an example of a half-Heusler alloy with a low figure of merit (zT) of just 0.05 \cite{yan2021effects}, despite having  power factor of 2.1 $\text{mW}/\text{mK}^{2} $  at room temperature. This low zT is primarily attributed to its lattice thermal conductivity, which has been experimentally measured to be 13.25 $\text{W}/\text{mK}$ \cite{yan2021effects}  and theoretically predicted to be 18$\text{W}/\text{mK} $ \cite{ye2021theoretical} at room temperature.
	  In this project, we investigated the thermoelectric properties of different structural
	  phases of $\text{Nb}_2\text{Co}_2\text{InSb}$ and $\text{Nb}_2\text{Co}_2\text{GaSb}$, which can be viewed as variants of NbCoSn where Sn atoms are replaced by In/Ga and Sb atoms. Our study included two
	  ordered structures and two Special Quasirandom Structures (SQSs). 
      The relative energetics analysis
	  reveals that, for $\text{Nb}_2\text{Co}_2\text{InSb}$, the ordered structure is the most stable one while for $\text{Nb}_2\text{Co}_2\text{GaSb}$, the SQS structure is the most stable configuration. Using the Debye-Callaway model \cite{asen1997thermal} \cite{zhang2012first} \cite{morelli2002estimation} , we calculated $k_{L}$ , finding it to range between 5.5 $\text{W}/\text{mK}$ and 6.9 $\text{W}/\text{mK}$ (4.7-5.8 $\text{W}/\text{mK}$) at room temperature for the different phases of  $\text{Nb}_2\text{Co}_2\text{InSb}$  $(Nb_{2}Co_{2}GaSb)$. We note that the values of $k_{L}$ for these double half-Heuslers are less than half of that observed in ternary NbCoSn.

	  \section{Introduction}
      Half-Heusler (hH) alloys represent a versatile class of intermetallic compounds with the general formula XYZ , where  X  and  Y  are  transition  metals, and  Z  is a p-block element. These materials crystallize in a non-centrosymmetric cubic structure (space group \( F\bar{4}3m \)) and have attracted significant attention due to their tunable electronic structures, mechanical robustness, and thermal stability. Their application in thermoelectric energy conversion stems from their semiconducting nature, high carrier mobility, and the possibility of band engineering through elemental substitution or disorder. This flexibility has led to the discovery of numerous high-performing thermoelectric compounds, such as ZrNiSn, HfNiSn, and TiNiSn \cite{gandi2016electron, zou2013electronic, sakurada2005effect}  for n-type transport as well as  NbFeSb, TaFeSb, and ZrCoBi \cite{naydenov2019huge, zhu2019discovery, zhu2018discovery}  for p-type transport. A key factor behind their efficiency lies in their outstanding electrical transport capabilities.
	  However, despite their promise, half-Heusler compounds face challenges when compared to leading thermoelectric materials like IV-VI group semiconductors. A major limitation is their inherently high lattice thermal conductivity ($k_{L}$), which hinders their performance. For instance, while ZrNiSn exhibits a relatively low $k_{L}$ of 10 Wm$^{\text{-1}}$K$^{\text{-1}}$\cite{zou2013electronic}  at room temperature—one of the best among half-Heusler materials—it remains significantly higher than the $k_{L}$ of state-of-the-art thermoelectrics like SnTe, and PbTe which has an intrinsic $k_{L}$ of just 2.5  Wm$^{\text{-1}}$K$^{\text{-1}}$ at 300 K \cite{xiao10charge, moshwan2017eco}. This discrepancy highlights the need for innovative approaches to design new materials that retain the desirable electronic properties of half-Heusler compounds while achieving much lower lattice thermal conductivity. \\

According to Zintl chemistry \cite{zeier2016engineering}, half-Heuslers (XYZ) are stable when the valence electron count (VEC) is 18, as this corresponds to fully filled bonding orbitals and empty anti-bonding orbitals. One effective approach to reducing the lattice thermal conductivity (\(k_L\)) is introducing mass disorder at lattice sites via aliovalent substitution, while maintaining a VEC of 18. This can be achieved by mixing VEC 17 and VEC 19 ternary half-Heusler compounds in equal proportions to form a quaternary double half-Heusler with VEC 18.

Shashwat Anand \textit{et al}.,~\cite{anand2019double} predicted 131 quaternary double half-Heuslers derived from ternary half-Heuslers using the \texttt{enumlib} code \cite{hart2008algorithm}, a significant increase compared to the previously identified 84 ternary half-Heusler systems. Among these, \( \text{Ti}_2\text{FeNiSb}_2 \), derived from \( \text{TiCoSb} \) (VEC 18) by combining \( \text{TiFeSb} \) (VEC 17) and \( \text{TiNiSb} \) (VEC 19) in equal amounts, was predicted to have a lattice thermal conductivity three times lower than the parent compound \( \text{TiCoSb} \).

Similarly, Bhawna Sahini \textit{et al}.,~\cite{sahni2024double} studied the thermoelectric performance of various structural phases of \( \text{Zr}_2\text{Ni}_2\text{InSb} \) and \( \text{Hf}_2\text{Ni}_2\text{InSb} \), which are derived from the ternary half-Heuslers \( \text{ZrNiSn} \) and \( \text{HfNiSn} \), respectively. Their findings indicate that these quaternary systems, particularly the ordered structures, exhibit better thermoelectric properties than their ternary counterparts.  \\

Like ZrNiSn and HfNiSn, NbCoSn is a half-Heusler semiconducting system with a band gap of 0.96 eV and a remarkably high melting temperature of approximately 2300 K \cite{wafula2022first}. Due to its thermal stability at high temperature, NbCoSn has been experimentally synthesized for thermoelectric applications \cite{yan2021effects}. Additionally, theoretical investigations into its thermoelectric properties have been reported \cite{ye2021theoretical}. The lattice thermal conductivity of NbCoSn has been estimated to be 18 Wm$^{\text{-1}}$K$^{\text{-1}}$ theoretically and 13.25 Wm$^{\text{-1}}$K$^{\text{-1}}$ experimentally. Despite possessing a power factor of 2.1 $\text{mW}$m$^{\text{-1}}$K$^{\text{-}2}$  at room temperature, its high lattice thermal conductivity limits its thermoelectric figure of merit ($zT$) to just 0.05 at room temperature.

This raises the question of whether derivative double half-Heusler structures can be developed from this parent compound, maintaining similar or better electronic properties while achieving significantly lower lattice thermal conductivity. In this project, we theoretically investigate the thermoelectric performance of $\text{Nb}_{2}\text{Co}_{2}\text{InSb}$ and $\text{Nb}_{2}\text{Co}_{2}\text{GaSb}$. These compounds can be viewed as derivative double half-Heusler structures derived from the parent half-Heusler compound NbCoSn, where the Sn atom is replaced by In/Ga and Sb atoms.

\begin{figure}[h!]
	  	\centering
	  	\includegraphics[width=\columnwidth]{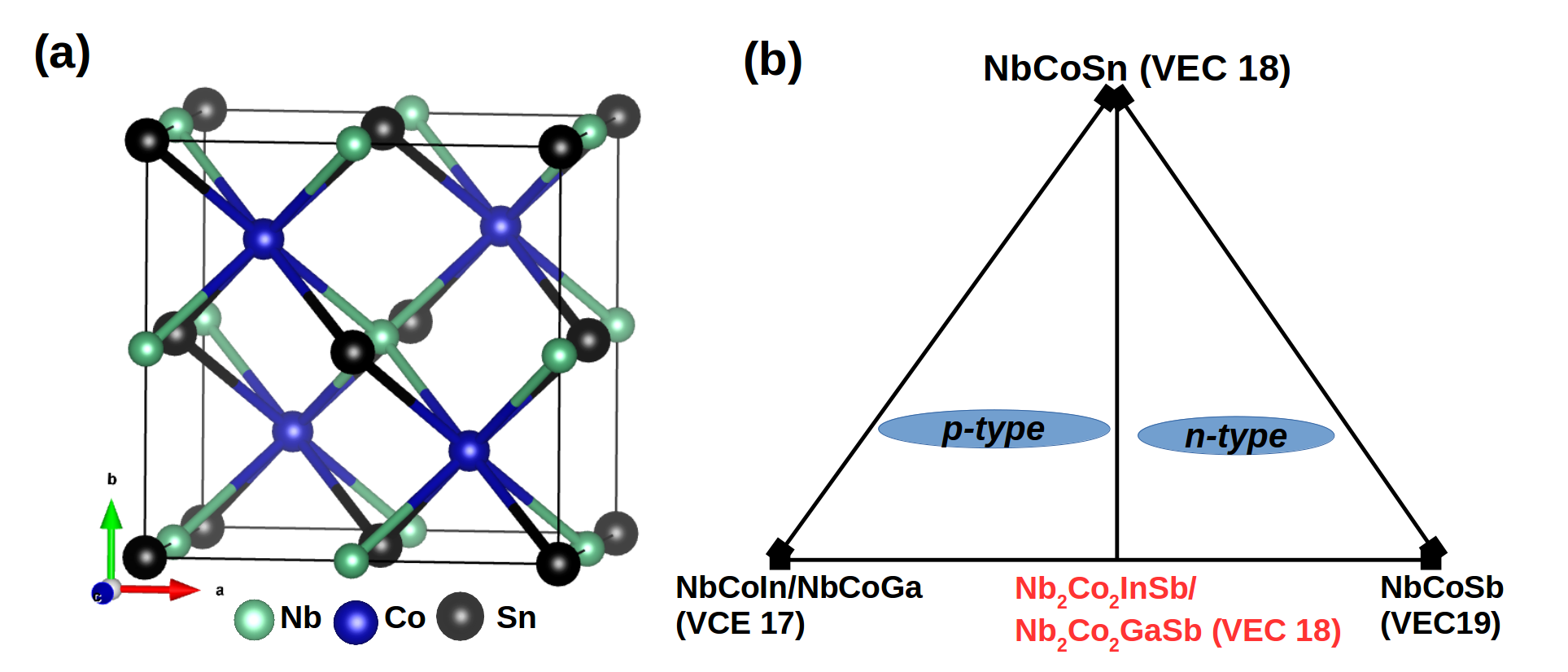} 
	  	\caption{(a) Crystal structure of NbCoSn (b) VEC 18 structural derivatives of NbCoSn }
	  	\label{fig:hh_dhh}
	  \end{figure}

As shown in Figure ~\ref{fig:hh_dhh}, these double half-Heusler alloys can be conceptualized as a combination of NbCoIn/NbCoGa ( VEC  17) and NbCoSb (VEC  19) structures in equal proportions. 
In this project, a comparative study of the thermoelectric properties of two ordered structures and two (one) disordered structures of $\text{Nb}_{2}\text{Co}_{2}\text{InSb}$ ($\text{Nb}_{2}\text{Co}_{2}\text{GaSb}$) was conducted. The results indicate that the ordered phases exhibit significantly superior electronic transport properties (power factor) compared to their disordered counterparts in both compounds. Furthermore, the lattice thermal conductivity of these double half-Heusler alloys is approximately five times lower than that of the ternary NbCoSn, making them highly promising candidates for thermoelectric applications.

	  \bigskip

      Unlike earlier studies, we have explored the properties of these compounds across  three relevant structural phases: cubic, tetragonal, and solid solution. For solid solution we generated two different Special Quasi-random structures SQSs (explained in section 3).  These phases can be realized depending on the synthesis conditions and temperature, with ordered phases forming at lower temperatures and disordered phases at higher temperatures. For our systems,  the ordered structures are  exhibiting higher carrier mobility compared to the solid solution phase. Combining electronic transport calculations with the thermal conductivity give better thermoelectric performance for the ordered structures for $\text{Nb}_2\text{Co}_2\text{GaSb}$ but for $\text{Nb}_2\text{Co}_2\text{InSb}$, the disorder structure has better thermoelectric performance.

 \section{Computational details}
	  Four different structural phases of both the systems were studied. Two of these are ordered structures and another two are Special Quasi-random structures(SQSs). The first ordered structure (OS1) was generated by replacing Sn atoms with In/Ga and Sb atoms in conventional unit cell. This generates Six different structures which was found to be energetically same after optimization. The second ordered structure (OS2) has been taken from Open Quantum Material Database (OQMD). OQMD predicts OS2 as the most stable structure. The SQSs were generated with a constraint that the In/Ga and Sb atoms
	  occupy the 4a(0, 0, 0) site. One of the SQSs is 12 atom SQSs without any constraint on the lattice structure (SQS1) while the second one is 24 atoms SQS with a contraint on the structure to be $2 \times 2 \times 2$ supercell of the primitive unit cell of half-Heusler structure (SQS2). In the construction of the SQS structures, pair correlation functions were included up to the fourth nearest-neighbor distance, while triplet correlations were considered up to the third nearest-neighbor distance. The SQS2 configuration of Nb$_{2}$Co$_{2}$GaSb was found to be dynamically unstable, and as a result, it was not considered for further analysis.

	  The calculations were carried out using plane-wave density functional theory (DFT) based calculations as implemented in the Quantum ESPRESSO \cite{giannozzi2017advanced, giannozzi2009quantum} software. The electron-ion interactions were described using ultrasoft pseudopotentials. For the wavefunction (charge density) we have used a basis set whose size corresponds to kinetic energy cutoff of 110 Ry (880 Ry). The electron-electron exchange and correlation effects were treated using the Perdew-Burke-Ernzerhof (PBE) \cite{perdew1996generalized} parametrization of the generalized gradient approximation (GGA). For electronic calculations, the Brillouin zone was sampled with a shifted $ 8\times 8\times 8$ for OS1, and OS2, $ 12\times 8\times 6$ for SQS1 and $ 4\times 4\times 4$ for SQS2 Monkhorst-pack k-mesh.  To compute the density of states (DOS) and the electronic transport properties, we  used $24 \times 24 \times 24$ k-mesh grid for OS1 and OS2, $ 36\times 24\times 18$ for SQS1 and $ 12\times 12\times 12$ for SQS2. To study the dynamical stability, vibrational properties and lattice thermal conductivity, the phonons were computed using density functional perturbation theory\cite{baroni2001phonons}. The calculations were performed on a q-mesh of $ 2\times 2\times 2$ for OS1, and OS2, $ 3\times 2\times 2$ for SQS1 and $ 2\times 2\times 2$ for SQS2. 
	  
	  Electronic transport properties were calculated by using the semi-classical 
	  Boltzmann transport theory within the constant relaxation time and rigid band 
	  approximations as implemented in the BoltzTrap2 code \cite{madsen2018boltztrap2}. Using BoltzTrap2 code Seebeck coefficient, electrical conductivity per unit relaxation time, thermal conductivity per unit relaxation time and power factor per unit relaxation time were calculated. The constant relaxation time was then calculated using deformation potential theory (as explained in section 2.7 of chapter 2) \cite{bardeen1950deformation}  and multiplied with the above transport coefficients per unit relaxation time to get the electronic transport coefficients.

	  \section{Results and Discussions}
	  
\begin{figure}[h!]
	  	\centering
	  	\includegraphics[width=\columnwidth]{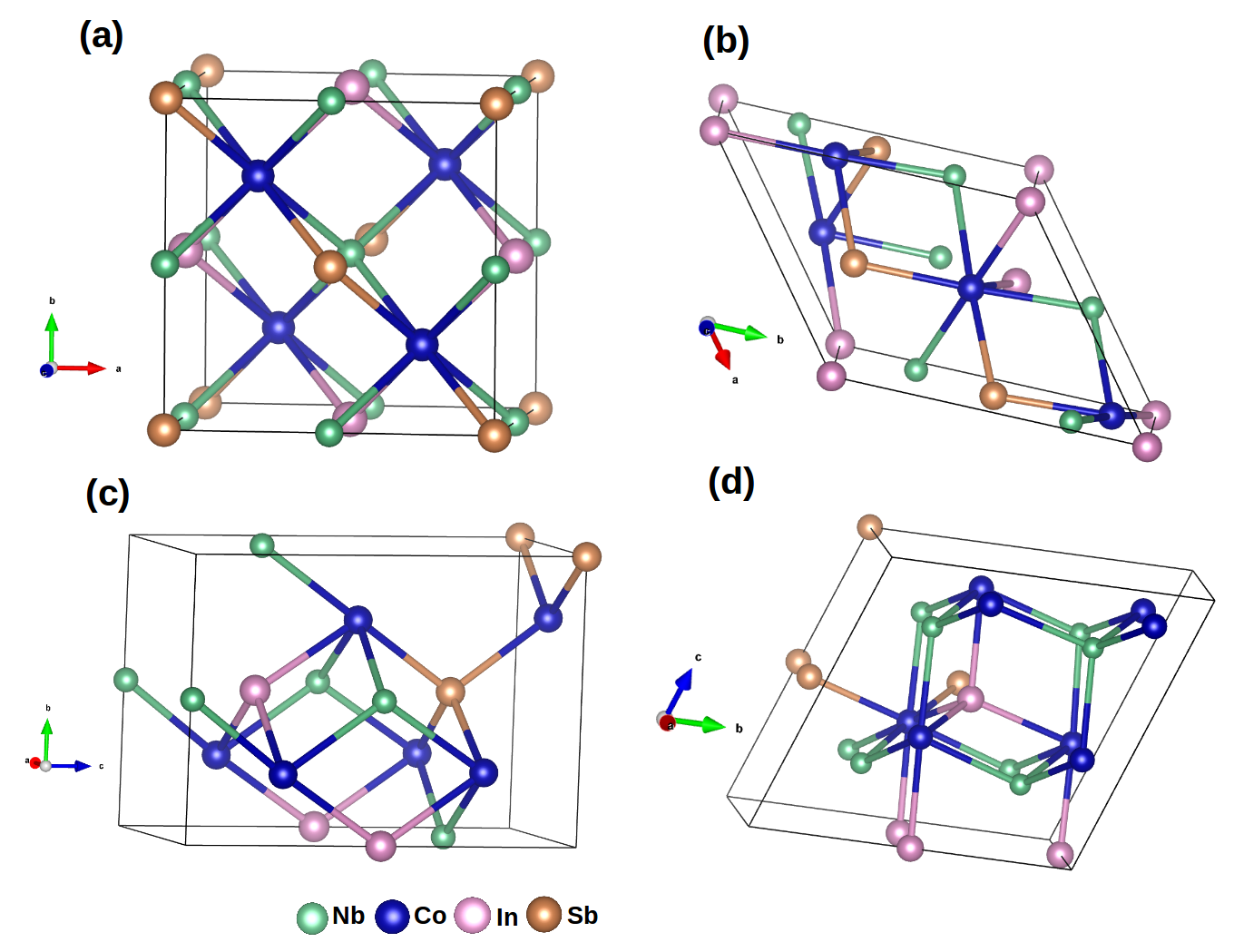} 
	  	\caption{Optimized crystal structure of $\text{Nb}_2\text{Co}_2\text{InSb}$ of (a) OS1 (b) OS2 (c) SQS1 (d) SQS2}
	  	\label{fig:insb_cs}
	  \end{figure}

\begin{figure}[h!]
	  	\centering
	  	\includegraphics[width=\columnwidth]{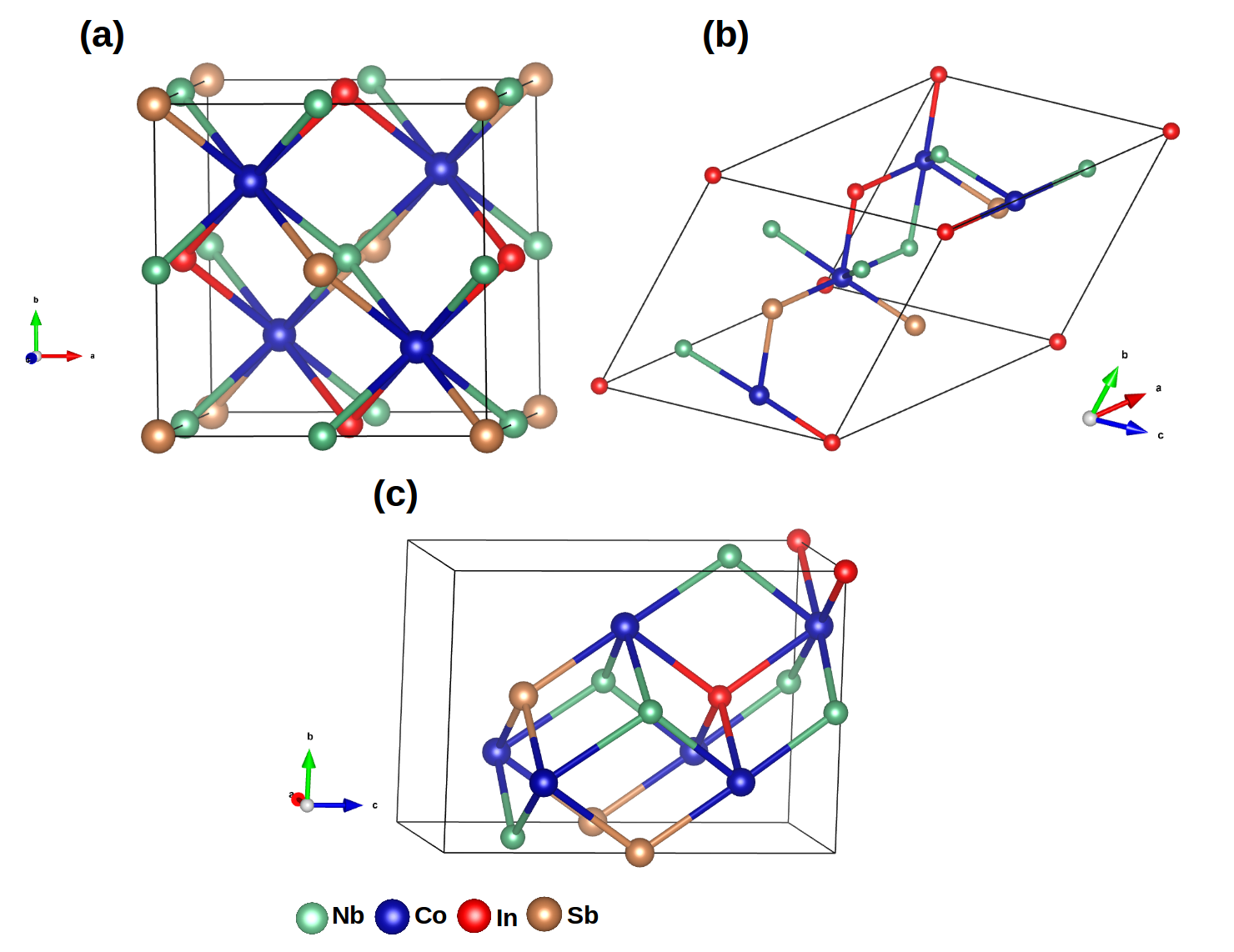} 
	  	\caption{Optimized crystal structure of $\text{Nb}_2\text{Co}_2\text{GaSb}$ of (a)OS1 (b) OS2 (c) SQS1 }
	  	\label{fig:gasb_cs}
	  \end{figure}

 The  crystal structures of all the structures are shown in the Figures~\ref{fig:insb_cs} and ~\ref{fig:gasb_cs} and their lattice parameters and relative energy is given in Tables~\ref{table:lp_insb} and ~\ref{table:lp_gasb}.

	  \bigskip

	  \bigskip

	  \setlength{\floatsep}{1em}
	  \begin{table} [!ht]
	  	\centering 
	  	\caption{Computed lattice parameters and the relative energies of all the structures of $\text{Nb}_2\text{Co}_2\text{InSb}$ .}
	  	\begin{tabular}{|c|c|c|c|}
	  		\hline
	  		Structure & \shortstack{Lattice \\parameters (\AA)} & Angles $(^\circ)$ & \shortstack{$\Delta E \ \text{per formula } $\\ unit (meV)} \\
	  		\hline 
	  		OS1 & \shortstack{a = b = 5.969,  \\ c = 5.962 } &  $\alpha $ = $\beta $ = $ \gamma = $  90 & 13.58 \\
	  		\hline
            OS2 & \shortstack{a = b = 5.964,  \\ c = 11.935 } & $\alpha $ = $\beta $ = $ \gamma = $  90 & 0 \\
            \hline
             SQS1 & \shortstack{a = 4.216,  b = 4.968,  \\ c = 8.439} & $\alpha $ = $\beta $ = $ \gamma = $  90 & 62.18 \\
              \hline
             SQS2 & \shortstack{a =   b = c = 8.438} & $\alpha $ = $\beta $ = $ \gamma = $  60 & 66.97 \\
             \hline
  
       	\end{tabular}
	  	\label{table:lp_insb}
	  \end{table} 
	  
	  \hspace{-40em}
	  
	  \begin{table} [!ht]
	  	\centering 
	  	\caption{Computed lattice parameters and the relative energies of all the structures of $\text{Nb}_2\text{Co}_2\text{GaSb}$ .}
	  	\begin{tabular}{|c|c|c|c|}
	  		\hline
	  		Structure & \shortstack{Lattice \\parameters (\AA)} & Angles $(^\circ)$ & \shortstack{$\Delta E \ \text{per formula } $\\ unit (meV)} \\
	  		\hline 
	  		OS1 & \shortstack{a = b = 5.859,  \\ c = 5.843 } &  $\alpha $ = $\beta $ = $ \gamma = $  90 & 115.51 \\
	  		\hline
	  		OS2 & \shortstack{a = b = 5.857,  \\ c = 11.684 } & $\alpha $ = $\beta $ = $ \gamma = $  90 & 12.06 \\
	  		\hline
	  		SQS1 & \shortstack{a = 4.121,  b = 5.867,  \\ c = 8.276} & $\alpha $ = $\beta $ = $ \gamma = $  90 & 0 \\
	  		\hline
	  			\end{tabular}
	  	\label{table:lp_gasb}
	  \end{table} 
	  
	  In Tables ~\ref{table:lp_insb} and ~\ref{table:lp_gasb}, $\Delta E$ is the change in energy per formula unit of a particular structure from the most stable structure.
	    As can be seen from the Tables ~\ref{table:lp_insb} that for $\text{Nb}_2\text{Co}_2\text{InSb}$, OS2 is the most stable structure (as predicted by OQMD) but for $\text{Nb}_2\text{Co}_2\text{GaSb}$,  SQS1 is the most stable configuration. Also for OS1, there is minor contraction of lattice parameter along the c axis changing it from cubic symmetry to tetragonal symmetry.
	    
	    \subsection{Electronic properties}
\begin{figure}[h!]
	    	  	\centering
	    	  	\includegraphics[width=\columnwidth]{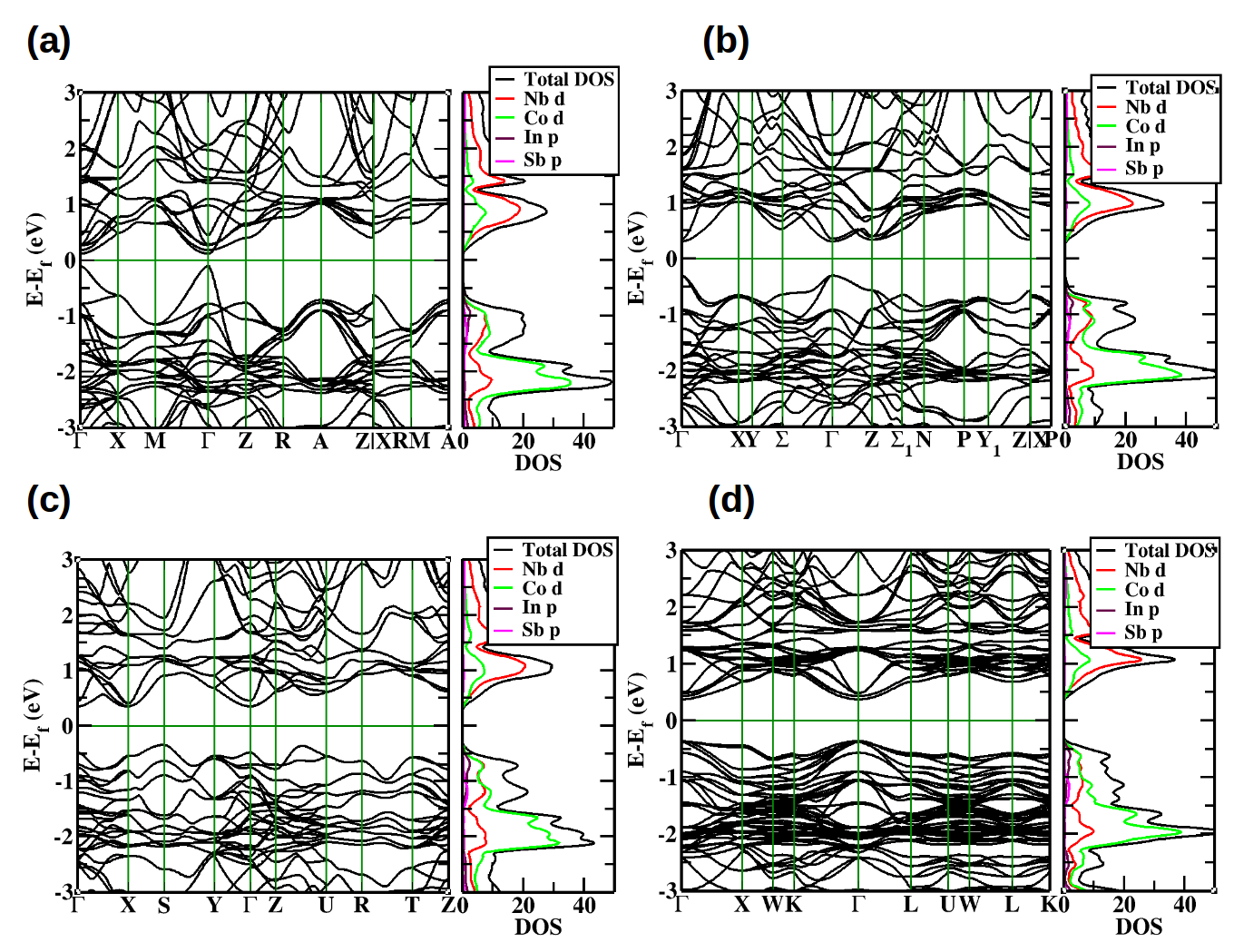} 
	    	  	\caption{Band structures and DOS of $\text{Nb}_2\text{Co}_2\text{InSb}$  (a)OS1 (b) OS2 (c) SQS1 (d) SQS2}
	    	  	\label{fig:insb_bs}
	    	  \end{figure}

\begin{figure}[h!]
	    	  	\centering
	    	  	\includegraphics[width=\columnwidth]{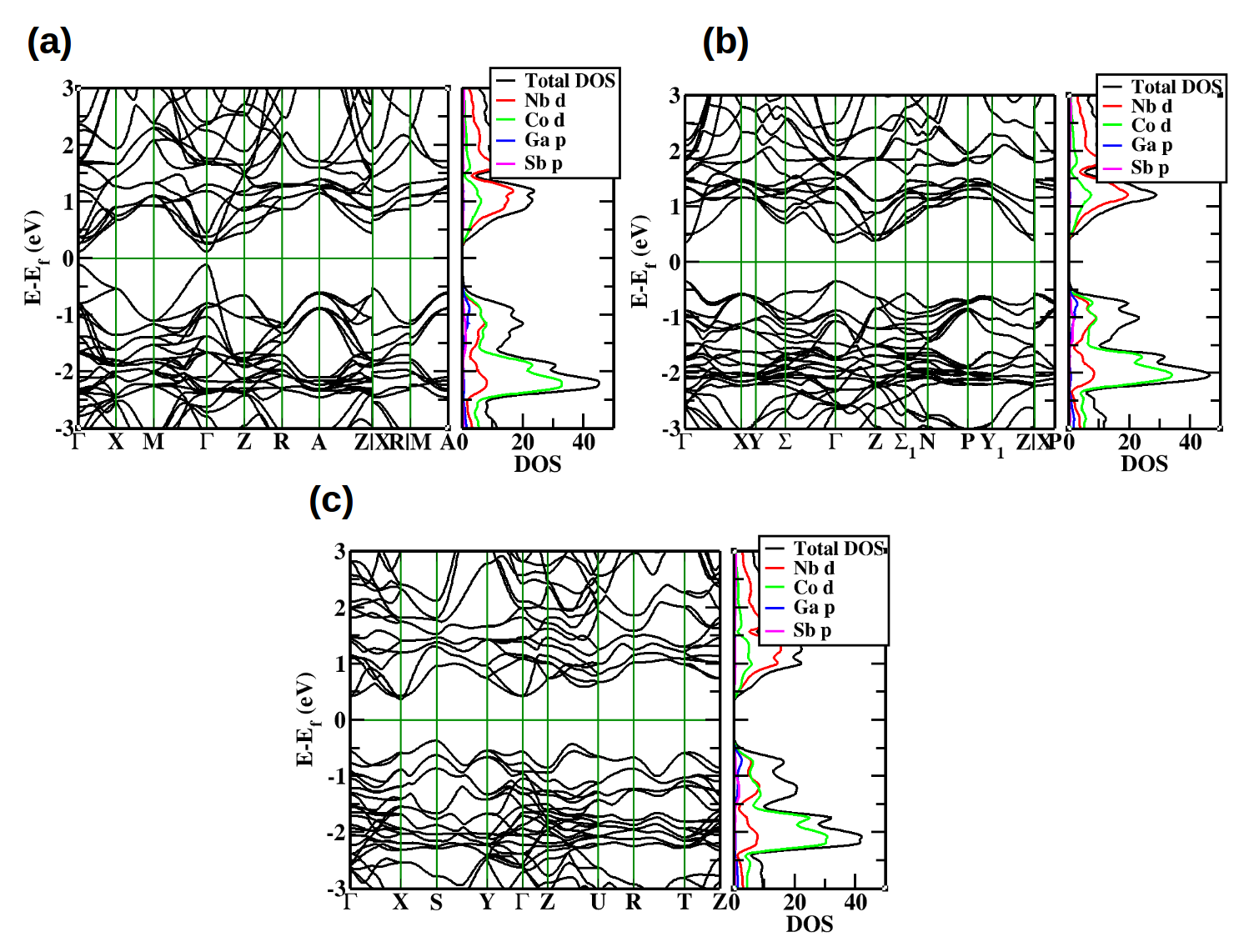} 
	    	  	\caption{Band structures and DOS of $\text{Nb}_2\text{Co}_2\text{GaSb}$  (a)OS1 (b) OS2 (c)SQS1 }
	    	  	\label{fig:gasb_bs}
	    	  \end{figure}

	     Figures~\ref{fig:insb_bs} and~\ref{fig:gasb_bs} show the band structure and density of states (DOS) of all the structures of $\text{Nb}_2\text{Co}_2\text{InSb}$ and $\text{Nb}_2\text{Co}_2\text{GaSb}$. As observed from the DOS plots, the valence and conduction band edges are primarily composed of Nb and Co $d$ orbitals, with a minor contribution from In/Ga $p$ orbitals. For both the systems, increasing disorder has led to an increse in the band gap OS1 being the lowest for both the systems and SQS2 being the highest for $\text{Nb}_2\text{Co}_2\text{InSb}$ and SQS1 for $\text{Nb}_2\text{Co}_2\text{GaSb}$. The band gap of the different structures are 0.21(0.23), 0.61(0.70), 0.69(0.74) and 0.74 eV for OS1, OS2, SQS1 and SQS2 respectively for  $\text{Nb}_2\text{Co}_2\text{InSb}$ ($\text{Nb}_2\text{Co}_2\text{GaSb}$). 
        In the OS1 structure for both systems, the valence band maximum (VBM) is located at the $\Gamma$ point. The conduction band minimum (CBM) also occurs at the $\Gamma$ point; however, an additional conduction band extremum (CBE) is present at the same point, lying 63~meV (130~meV) above the CBM for $\text{Nb}_2\text{Co}_2\text{InSb}$ ($\text{Nb}_2\text{Co}_2\text{GaSb}$).  
For the OS2 structure, the VBM again lies at the $\Gamma$ point for both systems. The CBM is also at $\Gamma$, while a CBE appears at the Z point, positioned 27~meV (28~meV) above the CBM for $\text{Nb}_2\text{Co}_2\text{InSb}$ ($\text{Nb}_2\text{Co}_2\text{GaSb}$).  
In the case of the SQS1 structure, the VBM is located at the S point, with another valence band extremum (VBE) situated between the Z and U points, which lies 18~meV (60~meV) below the VBM for $\text{Nb}_2\text{Co}_2\text{InSb}$ ($\text{Nb}_2\text{Co}_2\text{GaSb}$). The CBM for SQS1 is located at the X point in both systems. For $\text{Nb}_2\text{Co}_2\text{InSb}$, there is an additional CBM at the $\Gamma$ point, whereas in $\text{Nb}_2\text{Co}_2\text{GaSb}$, a CBE occurs at $\Gamma$, which is 53 meV above the CBM.  
In the SQS2 structure of $\text{Nb}_2\text{Co}_2\text{InSb}$, both the VBM and CBM are at the $\Gamma$ point. There is also a VBE at the same point lying 5 ~meV below the VBM and a CBE 58~meV above the CBM.

	    	  \bigskip

	    	  \bigskip








\subsection{Carrier Relaxation Times}
\label{sec:tau}
The average relaxation time ($\tau_{av}$) of charge carriers in these structures was determined using deformation potential theory\cite{bardeen1950deformation}, as detailed in Section 2.7 of Chapter 2. This method considers only the scattering of charge carriers by acoustic phonons. The conductivity and density of states effective masses at different band extrema in the Brillouin zone, which are required for calculating the relaxation time, are provided in Tables B.1 to B.8 in Appendix B.

	  \noindent The average relaxation time ($\tau_{av}$) was computed by considering all electronic bands situated within approximately 100~meV above the conduction band minimum (for electrons) and 100~meV below the valence band maximum (for holes).The parameters required to calculate $\tau_{av}$  along with the values of $\tau_{av}$ and $\mu_{av}$ for electrons and holes at 300 K for all structures, are presented in Tables~\ref{table:ele_ingasb} and \ref{table:hole_ingasb}, respectively. From these tables, it is evident that for $\text{Nb}_2\text{Co}_2\text{InSb}$, the holes have a lower effective mass compared to the electrons. For $\text{Nb}_2\text{Co}_2\text{GaSb}$, OS1 exhibits a lower effective mass for holes than electrons, while in OS2 and SQS1, the electrons have lighter bands than the holes.

\begin{figure}[h!]
		    	\centering
		    	\includegraphics[width=\columnwidth]{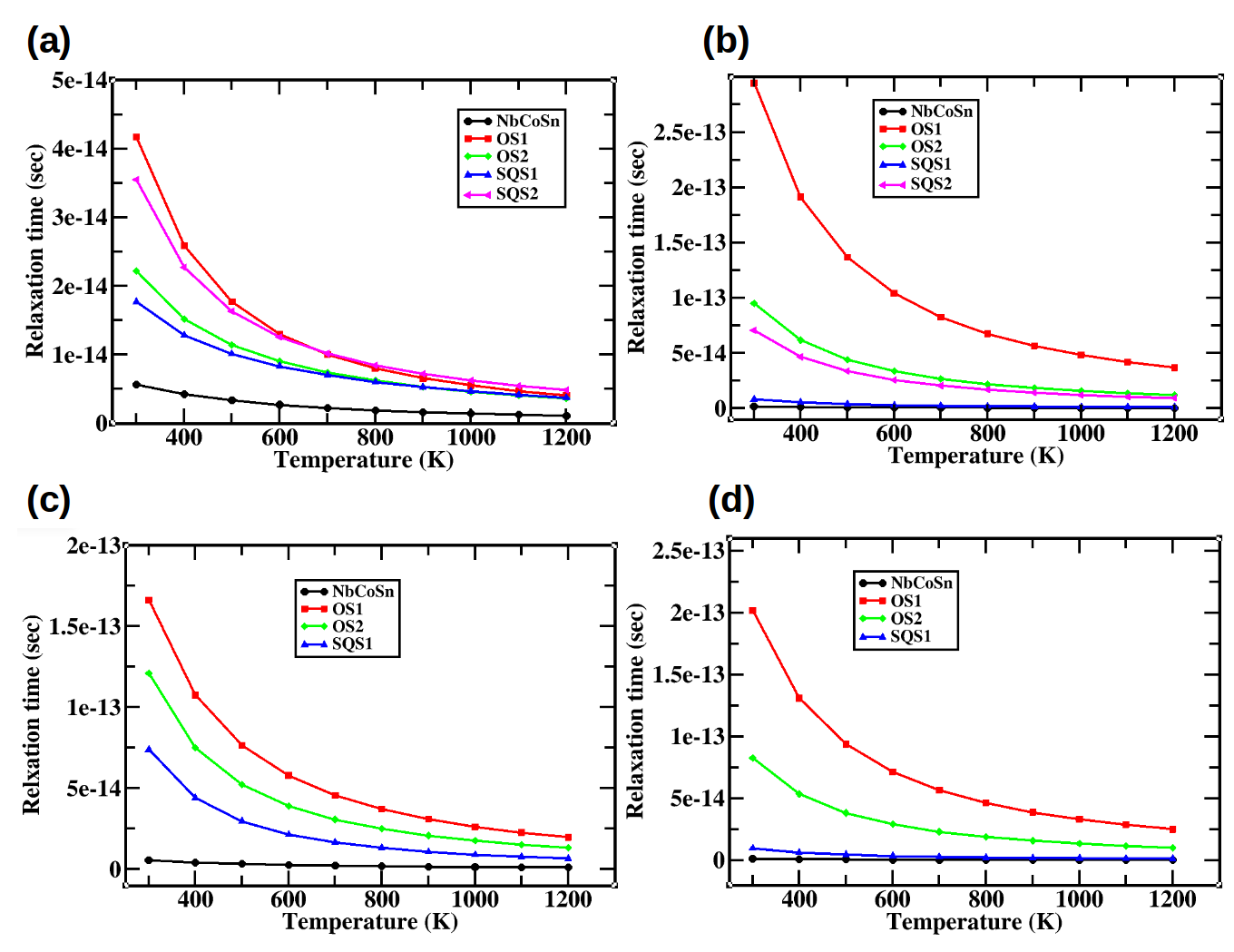} 
		    	\caption{ Carrier relaxation time for   (a) n-type $\text{Nb}_2\text{Co}_2\text{InSb}$ (b) p-type $\text{Nb}_2\text{Co}_2\text{InSb}$ (c) n-type $\text{Nb}_2\text{Co}_2\text{GaSb}$ (d) p-type $\text{Nb}_2\text{Co}_2\text{GaSb}$}
		    	\label{fig:tau_ingasb}
		    \end{figure}

\noindent In both systems, holes demonstrate exceptionally high carrier mobility ($\mu_{av}$) in the ordered structures, with the highest value of 1746.7 $\text{cm}^{2}/\text{Vs}$ observed in OS1 of $\text{Nb}_2\text{Co}_2\text{GaSb}$. For electrons, $\text{Nb}_2\text{Co}_2\text{GaSb}$ also shows remarkably high mobility across all structures. This high mobility for holes contributes to significantly greater relaxation times in OS1 and OS2, where the relaxation times are an order of magnitude higher than those of SQS1, as shown in the temperature-dependent relaxation time plots in Figure~\ref{fig:tau_ingasb}   (b) and (d).

\noindent Additionally, the deformation potential magnitudes for electrons and holes, which quantify the coupling strength between charge carriers and acoustic phonons, are found to be comparable.

 \begin{table} [ht]
	  	\centering 
	  	\caption{The average conductivity effective mass ($m_{\sigma, av}^{*}$), deformation potentials ($|\Xi|$), elastic constants ($C$), average mobility ($\mu_{av}$), and average relaxation time ($\tau_{av}$) for electrons in the various structural configurations of $\text{Nb}_2\text{Co}_2\text{InSb}$ ($\text{Nb}_2\text{Co}_2\text{GaSb}$) are presented. The values of $\mu_{av}$ and $\tau_{av}$ listed in the table were calculated at a temperature of 300 K.}
	  	\begin{tabular}{|c|c|c|c|c|c|}
	  		\hline
	  		Properties  & NbCoSn & OS1 & OS2 & SQS1 & SQS2   \\
	  		\hline 
	  		$m_{\sigma, av}^{*}$& 1.30 & 0.68 (0.33) & 1.49 (0.38) & 1.05 (0.47) & 0.76 \\
	  		\hline 
	  		$|\Xi|$ (eV) & 18.44 & 17.54 (17.49) & 17.48 (17.48) & 17.36 (17.42) & 17.41 \\
	  		\hline 
	  		$C$ (GPa) & 283.65 & 283.14 (292.01) & 323.56 (367.06) & 253.67 (276.97) & 276.67 \\
	  		\hline
	  		$\mu_{av}$ (cm$^{2}$/Vs) & 6.03 & 107.37 (890.3) & 26.20 (562.24) & 29.74 (278.43) & 80.49 \\
	  		\hline
	  		$\tau_{av}$ (fs) & 5.63 & 41.7 (166)  & 22.2(121) & 17.7 (73.9) &  35.5 \\	
	  		\hline
	  	\end{tabular}
	  	\label{table:ele_ingasb}
	  \end{table}
	  
	  \begin{table} [ht]
	  	\centering 
	  	\caption{The average conductivity effective mass ($m_{\sigma, av}^{*}$), deformation potentials ($|\Xi|$), elastic constants ($C$), average mobility ($\mu_{av}$), and average relaxation time ($\tau_{av}$) for holes in the various structural configurations of $\text{Nb}_2\text{Co}_2\text{InSb}$  ($\text{Nb}_2\text{Co}_2\text{GaSb}$) are presented. The values of $\mu_{av}$ and $\tau_{av}$ listed in the table were calculated at a temperature of 300 K.}
	  	\begin{tabular}{|c|c|c|c|c|c|}
	  		\hline
	  		Properties  & NbCoSn & OS1 &  OS2 & SQS1 & SQS2    \\
	  		\hline 
	  		$m_{\sigma, av}^{*}$ & 0.93 & 0.17 (0.20) & 0.49 (0.61) & 0.86 (0.79) & 1.93 \\
	  		\hline
	  		$|\Xi|$ (eV) & 17.65 & 17.29 (16.78) & 17.08 (17.03) & 16.75 (16.98) & 16.99 \\
	  		\hline
	  		$C$ (GPa)  & 283.65 & 283.14 (292.01) & 323.56 (367.06)& 253.67 (276.97) & 276.67 \\
	  		\hline
	  		$\mu_{av}$ (cm$^{2}$/Vs) & 2.91 & 297.98 (1746.7) & 339.69 (236.05) & 10.99(21.58) & 64.55 \\
	  		\hline
	  		$\tau_{av}$ (fs)     & 1.54 & 294 (202) & 95 (82.6) & 8.10 (9.73) & 70.7    \\
	  		\hline
	  	\end{tabular}
	  	\label{table:hole_ingasb}
	  \end{table}

		    \bigskip
		    
		    \subsection{Electronic transport calculations}

\begin{figure}[h!]
		   		    	\centering
		   		    	\includegraphics[width=\columnwidth]{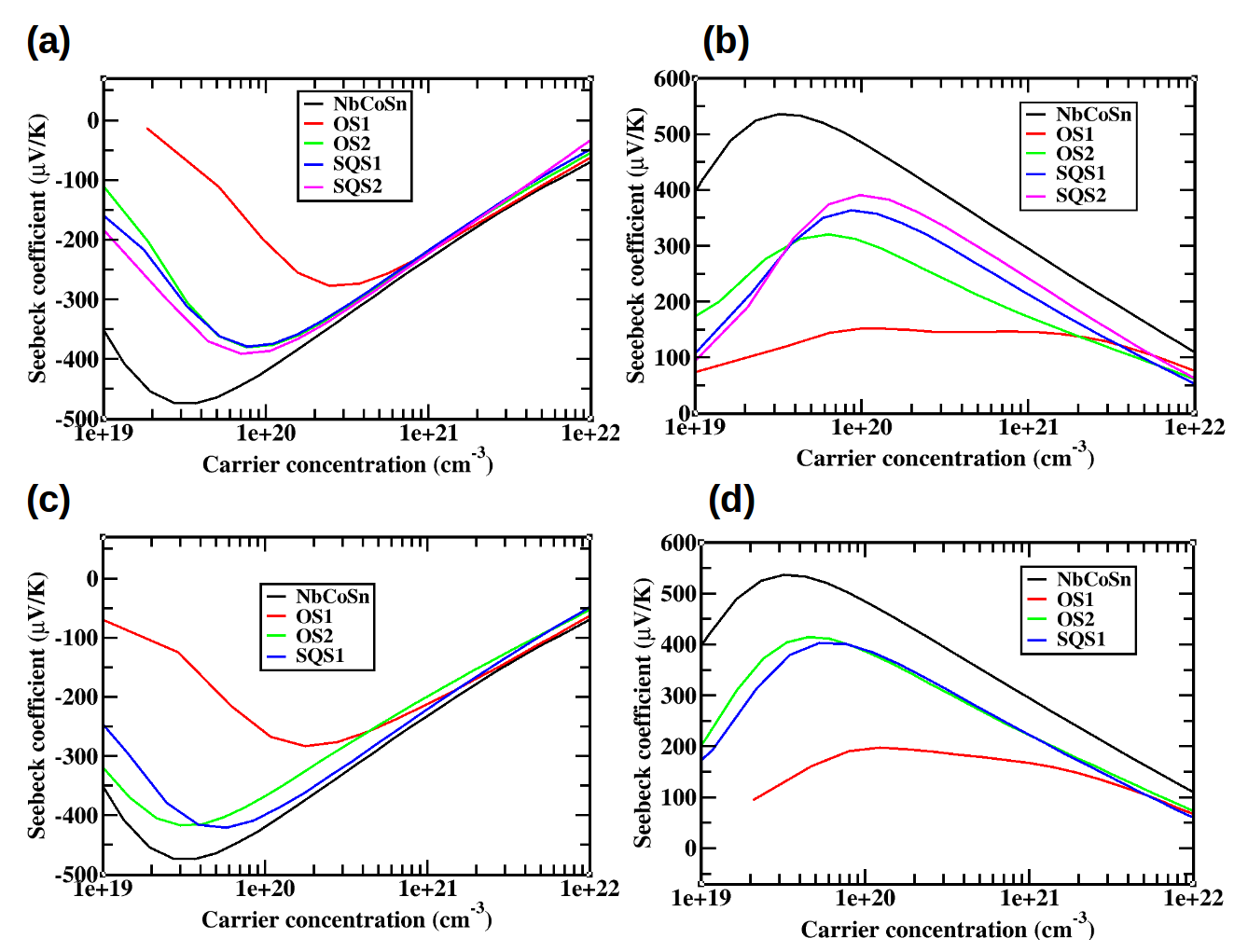} 
		   		    	\caption{ Seebeck coefficient at 1000K for  (a) n-type $\text{Nb}_2\text{Co}_2\text{InSb}$ (b) p-type $\text{Nb}_2\text{Co}_2\text{InSb}$ (c) n-type $\text{Nb}_2\text{Co}_2\text{GaSb}$ (d) p-type $\text{Nb}_2\text{Co}_2\text{GaSb}$}
		   		    	\label{fig:seebeck}
		   		    \end{figure}

\begin{figure}[h!]
		   		    	\centering
		   		    	\includegraphics[width=\columnwidth]{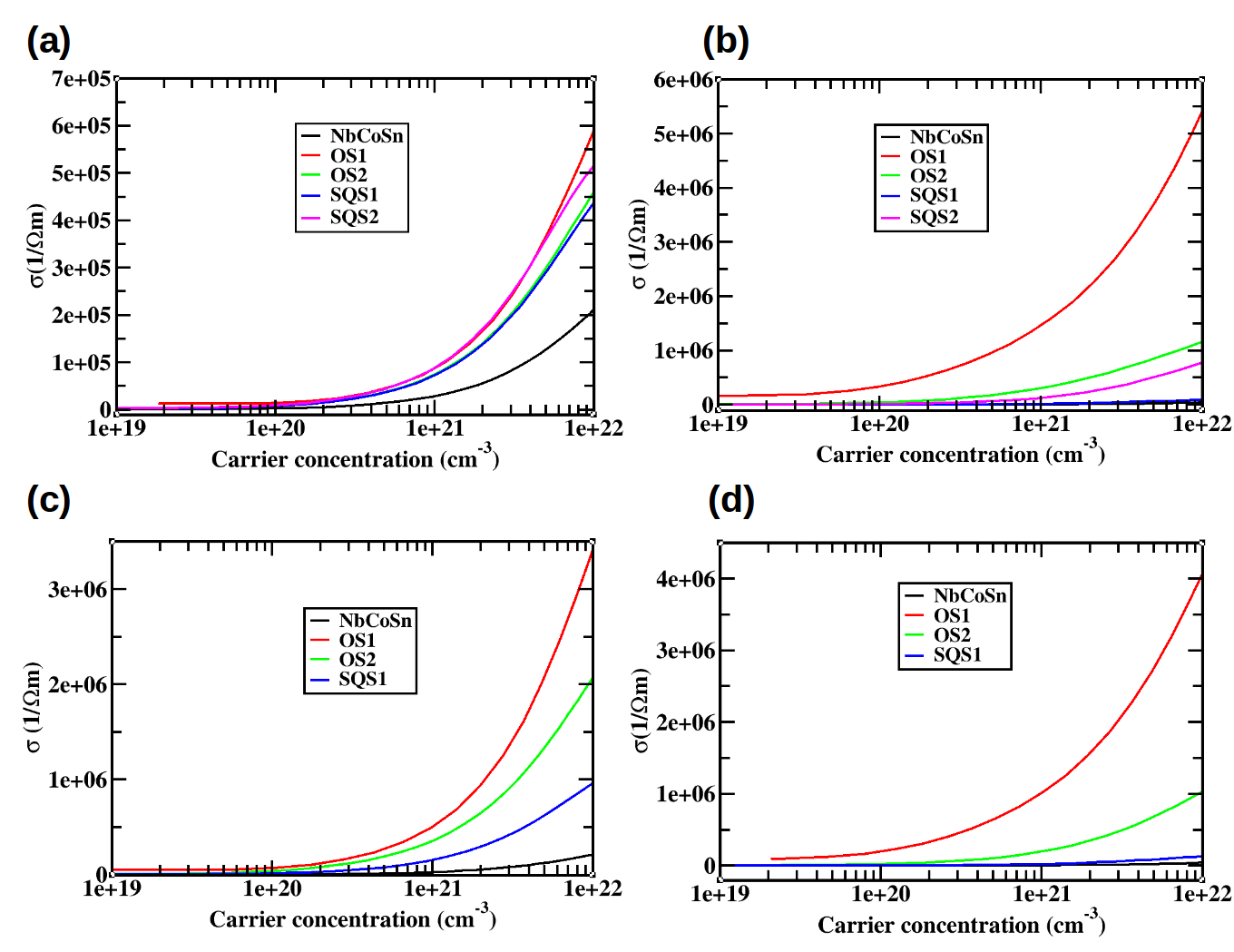} 
		   		    	\caption{Electrical conductivity at 1000K for  (a) n-type $\text{Nb}_2\text{Co}_2\text{InSb}$ (b) p-type $\text{Nb}_2\text{Co}_2\text{InSb}$ (c) n-type $\text{Nb}_2\text{Co}_2\text{GaSb}$ (d) p-type $\text{Nb}_2\text{Co}_2\text{GaSb}$}
		   		    	\label{fig:ec_ingasb}
		   		    \end{figure}

\begin{figure}[h!]
		   		    	\centering
		   		    	\includegraphics[width=\columnwidth]{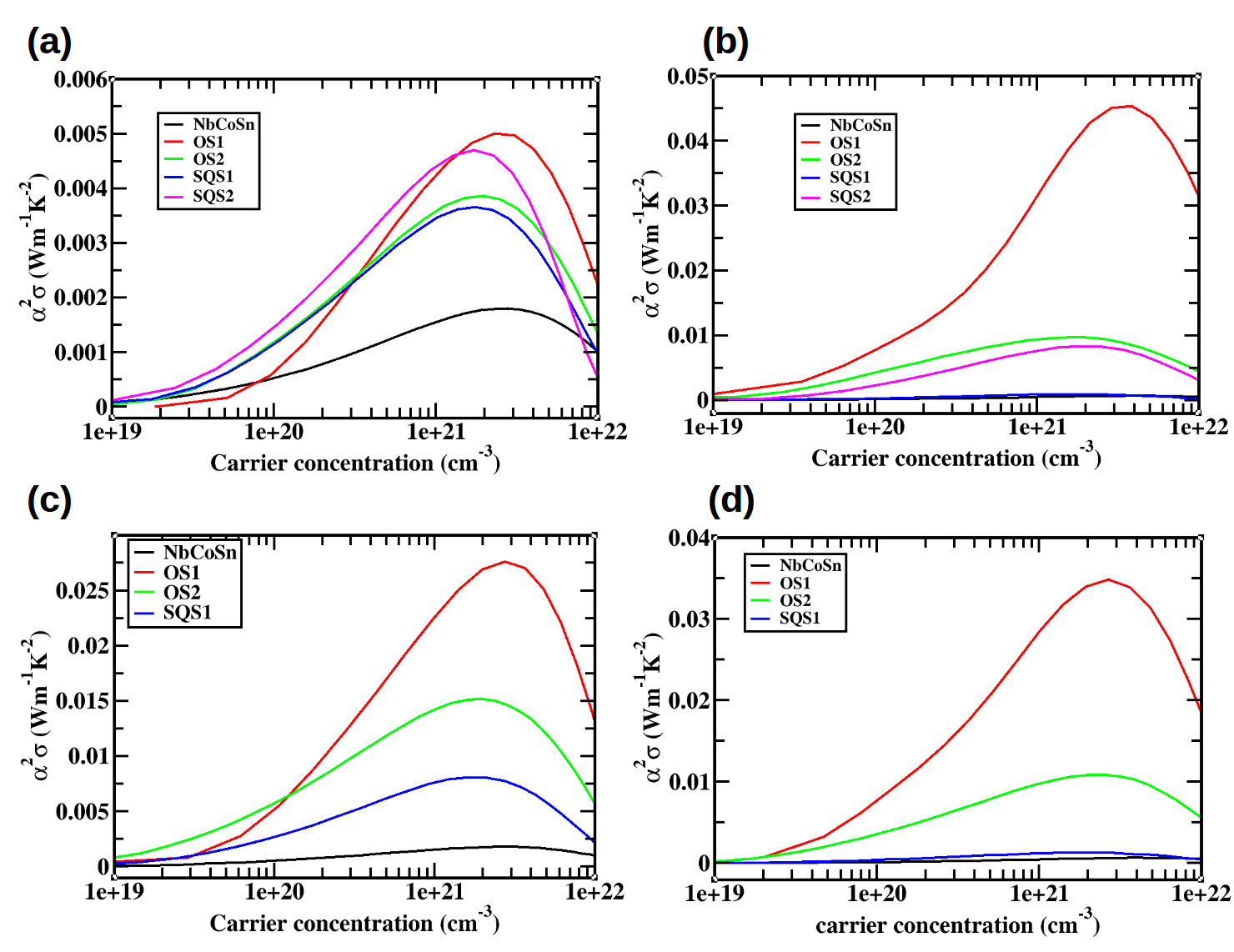} 
		   		    	\caption{ Power factor at 1000K for  (a) n-type $\text{Nb}_2\text{Co}_2\text{InSb}$ (b) p-type $\text{Nb}_2\text{Co}_2\text{InSb}$ (c) n-type $\text{Nb}_2\text{Co}_2\text{GaSb}$ (d) p-type $\text{Nb}_2\text{Co}_2\text{GaSb}$.}
		   		    	\label{fig:pf_ingasb}
		   		    \end{figure}

\noindent All electronic transport calculations were performed by combining the Boltzmann transport equation, assuming a constant relaxation time, with relaxation times derived from deformation potential theory. The Seebeck coefficient, electrical conductivity, and power factor as functions of carrier concentration at 1000 K are illustrated in Figures~\ref{fig:seebeck}, \ref{fig:ec_ingasb}, and \ref{fig:pf_ingasb}, respectively.\\
As shown in Figure~\ref{fig:seebeck}, the magnitude of the Seebeck coefficient for all the structures initially increases with carrier concentration, reaching a peak before decreasing. For NbCoSn in the n-type configuration, the maximum Seebeck coefficient is 473~$\mu$VK$^{\text{-}1}$ at a carrier concentration of $3.68 \times 10^{19}$~cm$^{\text{-}3}$, whereas in the p-type case, it peaks at 537~$\mu$VK$^{\text{-}1}$ at $3.19 \times 10^{19}$~cm$^{\text{-}3}$. 
For Nb$_2$Co$_2$InSb (n-type), the peak Seebeck coefficients for OS1, OS2, SQS1, and SQS2 are 277, 380, 378, and 391~$\mu$VK$^{\text{-}1}$, occurring at carrier concentrations of $2.46 \times 10^{20}$, $7.55 \times 10^{19}$, $7.69 \times 10^{19}$, and $6.98 \times 10^{19}$~cm$^{\text{-}3}$, respectively. For the p-type configuration of the same compound, the respective peak values are 151, 320, 364, and 391~$\mu$VK$^{\text{-}1}$ at $1.39 \times 10^{20}$, $6.34 \times 10^{19}$, $8.62 \times 10^{19}$, and $9.73 \times 10^{19}$~cm$^{\text{-}3}$. \\
In the case of Nb$_2$Co$_2$GaSb (n-type), OS1, OS2, and SQS1 exhibit peak Seebeck coefficients of 283, 417, and 421~$\mu$VK$^{\text{-}1}$ at carrier concentrations of $1.76 \times 10^{20}$, $3.01 \times 10^{19}$, and $5.78 \times 10^{19}$~cm$^{\text{-}3}$, respectively. For the p-type scenario, the peak values are 198, 415, and 403~$\mu$VK$^{\text{-}1}$ at $1.22 \times 10^{20}$, $4.51 \times 10^{19}$, and $5.24 \times 10^{19}$~cm$^{\text{-}3}$ for OS1, OS2, and SQS1, respectively.

Figure~\ref{fig:ec_ingasb} shows that the electrical conductivity of all the structures in both systems remains negligible up to a carrier concentration of approximately $1 \times 10^{20}$~cm$^{\text{-3}}$. This is because, at such low carrier concentrations, the chemical potential lies within the band gap, resulting in minimal electrical conductivity. As the carrier concentration increases, the conductivity rises accordingly.
Furthermore, from Figure~\ref{fig:ec_ingasb}(b) and (d), it is evident that for OS1, the electrical conductivity in the p-type case is about an order of magnitude higher than that of the other structures for both systems. This behavior is consistent with the values presented in Table~\ref{table:hole_ingasb}, which shows that OS1 has a lower average effective mass compared to the other configurations. A similar trend is observed for the n-type case, where OS1 also exhibits higher electrical conductivity due to its lower average effective mass, as listed in Table~\ref{table:ele_ingasb}. 

As observed in Figures~\ref{fig:seebeck} and \ref{fig:ec_ingasb}, the magnitude of the Seebeck coefficient decreases while the electrical conductivity increases with increasing carrier concentration beyond approximately $1 \times 10^{20}$~cm$^{\text{-3}}$. This complementary behavior leads to an optimal value of the power factor, $\alpha^{2}\sigma$, as illustrated in Figure~\ref{fig:pf_ingasb}. Particularly for the p-type configuration, OS1 exhibits a significantly higher peak power factor compared to the other structures.
For NbCoSn, the peak power factor in the n-type case is 1.80~mWm$^{\text{-1}}$K$^{\text{-2}}$, occurring at a carrier concentration of $2.76 \times 10^{21}$~cm$^{\text{-3}}$, while for the p-type case, the maximum value is 0.66~mWm$^{\text{-1}}$K$^{\text{-2}}$ at $4.35 \times 10^{21}$~cm$^{\text{-3}}$.
In the case of Nb$_2$Co$_2$InSb (n-type), the peak power factors for OS1, OS2, SQS1, and SQS2 are 5.01, 3.87, 3.66, and 4.70~mWm$^{\text{-1}}$K$^{\text{-2}}$ at carrier concentrations of $2.28 \times 10^{21}$, $1.94 \times 10^{21}$, $1.75 \times 10^{21}$, and $1.72 \times 10^{21}$~cm$^{\text{-3}}$, respectively. For the p-type case, the corresponding peak values are 45.40, 9.72, 0.91, and 8.38~mWm$^{\text{-1}}$K$^{\text{-2}}$ at carrier concentrations of $3.84 \times 10^{21}$, $1.53 \times 10^{21}$, $1.59 \times 10^{21}$, and $1.86 \times 10^{21}$~cm$^{\text{-3}}$.
For Nb$_2$Co$_2$GaSb in the n-type case, the maximum power factors for OS1, OS2, and SQS1 are 27.62, 15.18, and 8.11~mWm$^{\text{-1}}$K$^{\text{-2}}$ at carrier concentrations of $2.73 \times 10^{21}$, $1.93 \times 10^{21}$, and $1.64 \times 10^{21}$~cm$^{\text{-3}}$, respectively. For the p-type case, OS1, OS2, and SQS1 reach peak power factors of 34.85, 10.91, and 1.30~mWm$^{\text{-1}}$K$^{\text{-2}}$ at carrier concentrations of $2.66 \times 10^{21}$, $2.46 \times 10^{21}$, and $2.20 \times 10^{21}$~cm$^{\text{-3}}$, respectively.

		   		    \bigskip

	  \subsection{Phonon dispersion and Lattice thermal conductivity}

\begin{figure}
		\centering
		\includegraphics[scale=0.40]{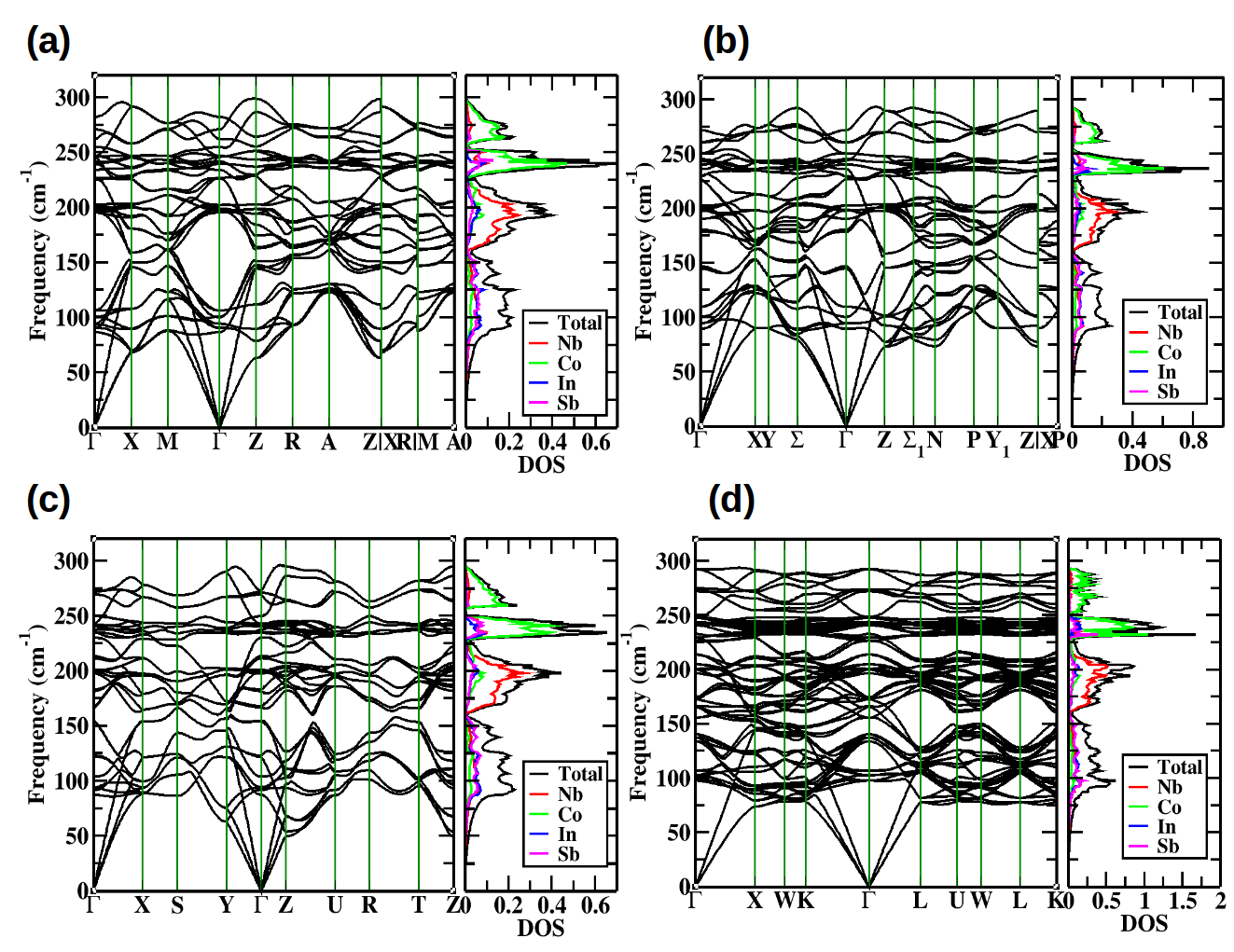}
		\caption{Phonon dispersions and phonon DOS of different structures of $\text{Nb}_2\text{Co}_2\text{InSb}$ (a) OS1 (b) OS2 (c) SQS1 (d) SQS2.}
		\label{fig:phonon_insb}
	\end{figure}

\begin{figure}
		\centering
		\includegraphics[scale=0.40]{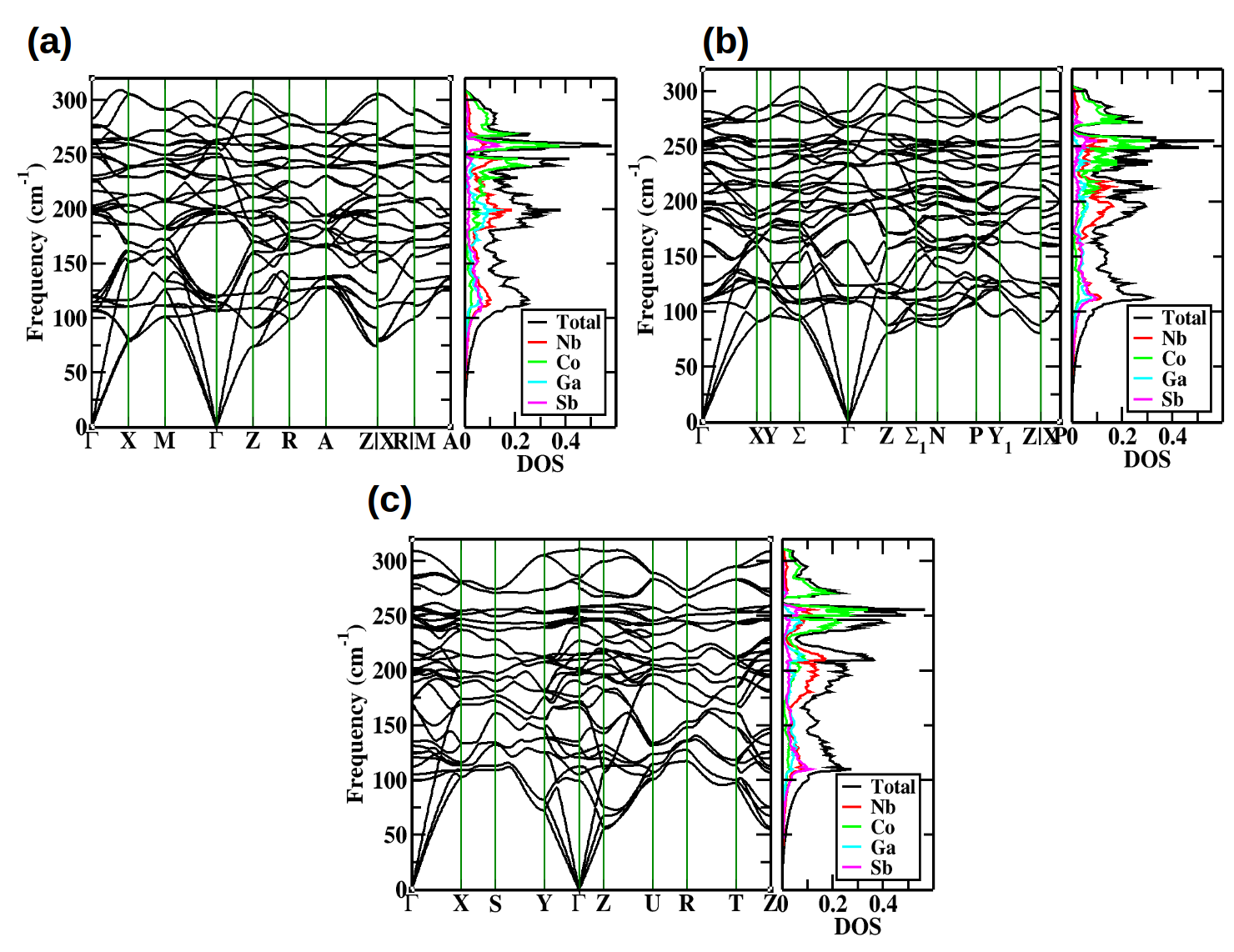}
		\caption{Phonon dispersions and phonon DOS of different structures of $\text{Nb}_2\text{Co}_2\text{GaSb}$ (a) OS1 (b) OS2 (c) SQS1. }
		\label{fig:phonon_gasb}
	\end{figure}

\noindent To assess the dynamical stability of both systems, phonon dispersion curves were calculated for all structures of $\text{Nb}_2\text{Co}_2\text{InSb}$ and $\text{Nb}_2\text{Co}_2\text{GaSb}$. The phonon dispersion curves are presented in Figure~\ref{fig:phonon_insb} for $\text{Nb}_2\text{Co}_2\text{InSb}$ and Figure ~\ref{fig:phonon_gasb} for $\text{Nb}_2\text{Co}_2\text{GaSb}$. Among all the considered structures, only SQS2 of Nb$_2$Co$_2$GaSb exhibited dynamical instability due to the presence of imaginary phonon modes. The phonon dispersion curve of SQS2 of Nb$_2$Co$_2$GaSb is shown in Figure B.5 of Appendix B.   In contrast, all other configurations for both systems showed no imaginary frequencies, indicating their dynamical stability.
 \noindent From the phonon density of states (PDOS), it is evident that the low-frequency range (0–150 cm$^{\text{-}1}$) is primarily contributed by Sb, In/Ga, and Nb atoms. In the intermediate range (150–230 cm$^{\text{-}1}$), Nb atoms dominate, whereas the higher frequency range (above 230 cm$^{\text{-}1}$) is predominantly influenced by the lighter Co atoms.
	
To assess anharmonicity in the structures, the mode-resolved Grüneisen parameter ($\gamma_i^k$) was calculated. For a phonon in the $i^{th}$ branch with wave vector $k$, it is defined as:

\begin{equation}
    \gamma_{i}^{k} = \text{-}\frac{V_0}{\omega_{i}^{k}}\frac{\partial \omega_{i}^{k}}{\partial V}
    \label{eq:gru_mode}
\end{equation}

\noindent where $V_0$ is the equilibrium volume of the unit cell, and $\omega_{i}^{k}$ represents the frequency of the corresponding phonon. The derivative  in Equation~\ref{eq:gru_mode} was computed numerically using the central difference method.  For all the dynamically stable structures except SQS1 of $\text{Nb}_2\text{Co}_2\text{GaSb}$, the Gruneisen parameters were calculated using the phonon dispersions by increasing and decreasing the lattice parameters by 1\%. In the case of SQS1 of $\text{Nb}_2\text{Co}_2\text{GaSb}$, a smaller change of 0.5\% was used instead. This was because a 1\% increment led to bending of the transverse acoustic modes near the $\Gamma$ point, resulting in a non-linear dispersion curve in that region.

\begin{figure}
	\centering
	\includegraphics[scale=0.35]{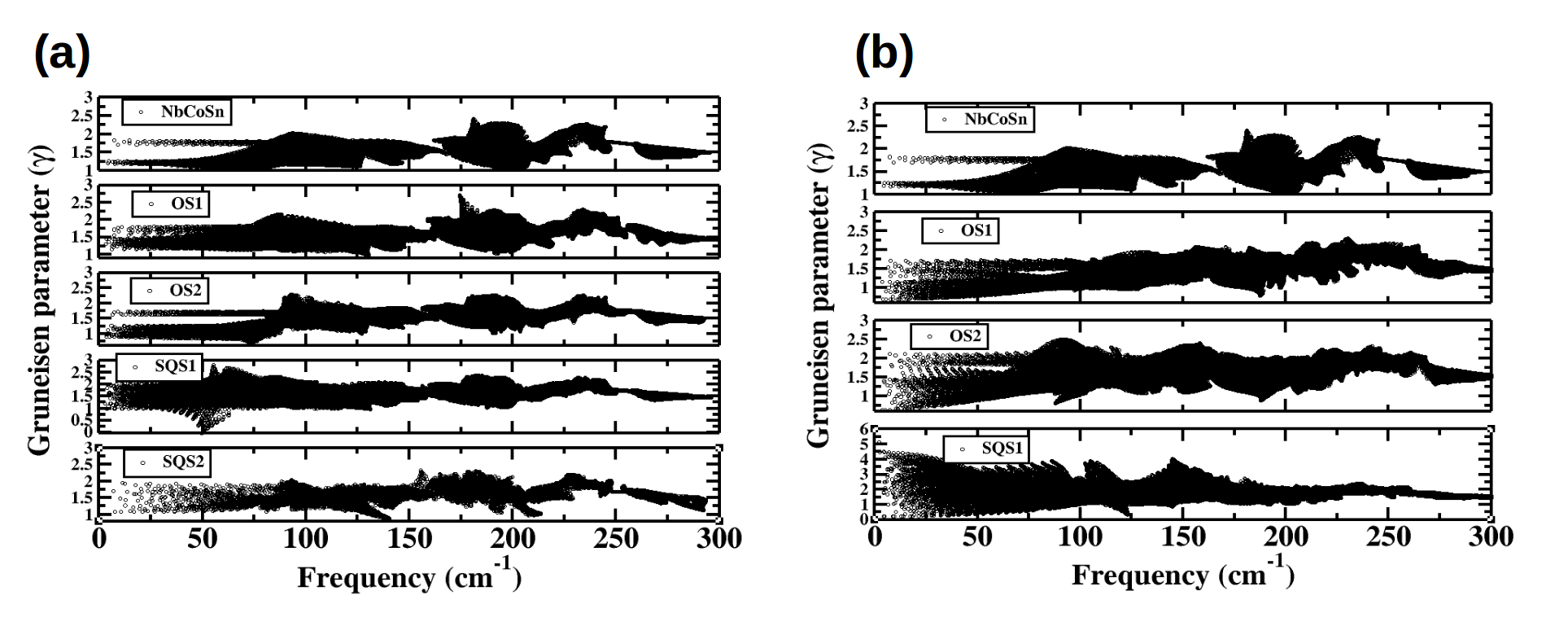}
	\caption{Mode resolved Gruneisen parameters for different structures of (a) $\text{Nb}_2\text{Co}_2\text{InSb}$ and (b) $\text{Nb}_2\text{Co}_2\text{GaSb}$.}
	\label{fig:gru_ingasb}
\end{figure}

Figure~\ref{fig:gru_ingasb} illustrates the Grüneisen parameters for different structures of both $\text{Nb}_2\text{Co}_2\text{InSb}$ and $\text{Nb}_2\text{Co}_2\text{GaSb}$. The results indicate that SQS1 exhibits greater anharmonicity in both systems. Its Grüneisen parameters span a wider range, varying from 0 to 3 for $\text{Nb}_2\text{Co}_2\text{InSb}$ and 0 to 6 for $\text{Nb}_2\text{Co}_2\text{GaSb}$, compared to the narrower range of 0.5 to 2.5 observed in other structures.

The lattice thermal conductivity ($k_L$) for all systems was calculated using the Debye-Callaway model \cite{asen1997thermal, zhang2012first, morelli2002estimation} . This model considers $k_L$ as the sum of contributions from the longitudinal acoustic ($k_{LA}$) and two transverse acoustic branches ($k_{TA}$ and $k_{TA'}$). The methodology for calculating the lattice thermal conductivity is detailed in section 2.8 of chapter 2. \\

In most crystalline solids, resistive scattering arises primarily due to Umklapp processes. Hence, for half-Heusler NbCoSn, the total phonon scattering rate for $i^{th}$ mode combines the contributions from normal and Umklapp processes as:

\begin{equation}
    \frac{1}{\tau_{HH,i}^{C}} = \frac{1}{\tau_{i}^{N}} + \frac{1}{\tau_{i}^{U}}
    \label{n_um}
\end{equation}

However, in the double Heusler compounds $\text{Nb}_2\text{Co}_2\text{InSb}$ and $\text{Nb}_2\text{Co}_2\text{GaSb}$, significant mass and strain field fluctuations occur due to the Wyckoff position 4c (0.0, 0.0, 0.0) being occupied by either In/Ga or Sb atoms, which differ in mass and atomic radius. This results in additional phonon scattering from these point defects.       Therefore, in the double half-Heuslers (DHH), propagating phonons are also expected to be scattered by mass and strain field defects arising from these fluctuations, which play a significant role in phonon scattering. Hence, to compute the
scattering rates for dissipative processes in the double half-Heuslers, we have also 
incorporated the effect of mass and strain field fluctuation scattering. Consequently, the total relaxation time for the $i^{th}$ mode of DHH is given by : \\
\begin{equation}
\frac{1}{\tau_{DHH,i}^{C}} = \frac{1}{\tau_{i}^{N}} \ + \frac{1}{\tau_{i}^{U}} + \frac{1}{\tau_{i}^{M}} + \frac{1}{\tau_{i}^{S}}
\label{eq:tau_full_hea}
\end{equation}

The parameters used to compute $k_L$  are summarized in Table~\ref{table:pho_para}.

	  \begin{table} [ht]
	  	\centering 
	  	\caption{ The longitudinal velocity, transverse velocities, mode-resolved Debye temperature ($\theta_{i}$), mode-resolved Gruneisen parameters ($\gamma_{i}$), and the total Gruneisen parameter ($\gamma$) for NbCoSn and the different structures of $\text{Nb}_2\text{Co}_2\text{InSb}$ ($\text{Nb}_2\text{Co}_2\text{GaSb}$) are listed below.}
	  	\begin{tabular}{|c|c|c|c|c|c|}
	  		\hline
             System & \shortstack{Longitudinal \\ velocity $(v_{LA}$) \\ (m/sec)} & \shortstack{Transverse \\velocity \\ ($v_{TA}$, $v_{TA^{'}}$) \\ (m/sec)} & \shortstack{ Debye \\ temperatures \\ $\theta_{LA} , \  \theta_{TA}, \  \theta_{TA^{'}}$  (K)} & \shortstack{mode \\ Gruneisen \\parameters \\ $\gamma_{LA}, \gamma_{TA}, \gamma_{TA^{'}}$} & \shortstack{Total\\Gruneisen \\ parameter \\($\gamma$)} \\
             \hline
             NbCoSn & 5852 & 2912, 2912 & 242, 182, 184 & 1.71, 1.52, 1.46 & 1.67 \\ 
            
             \hline 
                OS1 & \shortstack{6109 \\ (6315)} & \shortstack{2842, 2902 \\ (3150, 3265)} & \shortstack{179, 176, 178 \\ (185, 183, 184)}  & \shortstack{1.57, 1.33, 1.35 \\ (1.30, 1.05, 1.09)}  & \shortstack{1.67 \\ (1.61)} \\
            \hline
            
             OS2 & \shortstack{5788 \\ (6087)} & \shortstack{2869, 3038 \\ (3038, 3399)} & \shortstack{176, 153, 155 \\ (176, 155, 160)} &  \shortstack{1.58, 1.24, 1.37 \\ (1.55, 1.46, 1.45)} & \shortstack{1.68 \\ (1.71)} \\
             \hline
             
             SQS1 & \shortstack{5470 \\ (5382)} & \shortstack{2713, 3403 \\ (2570, 3848)} & \shortstack{161, 146, 147 \\ (183, 168, 168) }& \shortstack{1.67, 1.64, 1.64 \\ (1.70, 1.91, 1.54)} & \shortstack{1.70 \\ (1.75)} \\
             \hline
             SQS2 & 5290 & 3223, 3223  & 135, 113, 121 & 1.53, 1.31, 1.48 & 1.68 \\
             \hline
          \end{tabular}
          	  	\label{table:pho_para}
	  	 \end{table}

\begin{figure}
	\centering
	\includegraphics[scale=0.35]{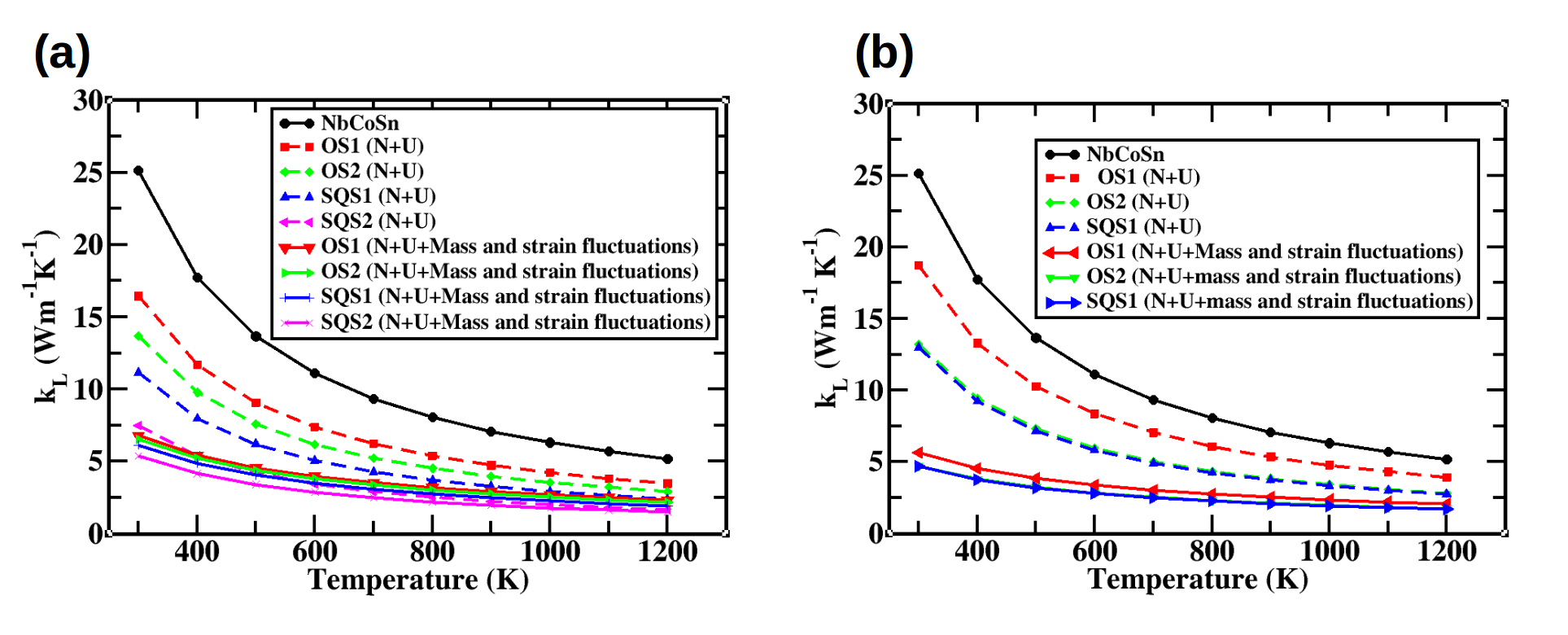}
	\caption{Lattice thermal conductivity as a function of temperature for different structural phases of (a) $\text{Nb}_2\text{Co}_2\text{InSb}$ and (b) $\text{Nb}_2\text{Co}_2\text{GaSb}$.}
	\label{fig:kappal_ingasb}
\end{figure}

\noindent Figure~\ref{fig:kappal_ingasb} illustrates the lattice thermal conductivity of various structures for both systems. As evident from the plots, the lattice thermal conductivity at a given temperature decreases with increasing disorder in the system. For $\text{Nb}_2\text{Co}_2\text{InSb}$ at 300~K, the ordered structure OS1 exhibits the highest lattice thermal conductivity of 16.45 (6.78) $\, \text{Wm}^{\text{-}1}\text{K}^{\text{-}1}$ without (with) mass and strain fluctuation scattering, while the disordered structure SQS2 shows the lowest values of 7.49 (5.40) $\, \text{Wm}^{\text{-}1}\text{K}^{\text{-}1}$ without (with) mass and strain field fluctuation scattering. Similarly, for $\text{Nb}_2\text{Co}_2\text{GaSb}$ at 300~K, OS1 has the highest lattice thermal conductivity of 18.71 (5.66)$\, \text{Wm}^{\text{-}1}\text{K}^{\text{-}1}$ without (with) mass and strain field fluctuation scattering, whereas SQS1 has the lowest values of 12.99 (4.70)$\, \text{Wm}^{\text{-}1}\text{K}^{\text{-}1}$ without(with) mass and strain field fluctuation scattering. Notably, the lattice thermal conductivity of these double half-Heusler compounds is approximately one-fifth that of the corresponding ternary half-Heusler compound, $\text{NbCoSn}$.


 \subsection{Figure of merit}
 
\begin{figure}[h!]
 	\centering
 	\includegraphics[width=\columnwidth]{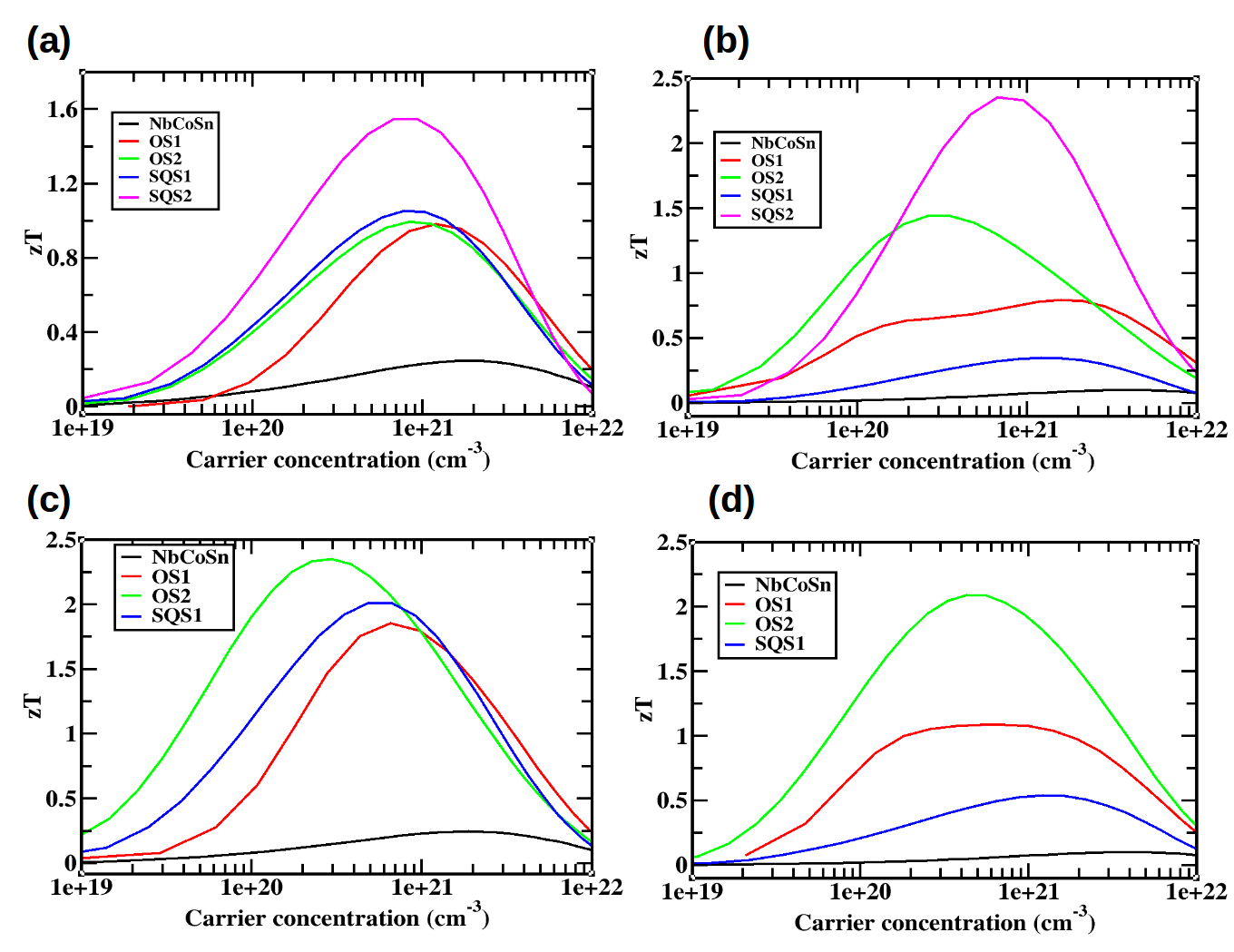} 
 	\caption{ Figure pf merit $zT$ at 1000 K as a function of carrier concentration for (a) n-type $\text{Nb}_2\text{Co}_2\text{InSb}$, (b)  p-type $\text{Nb}_2\text{Co}_2\text{InSb}$ , (c)  n-type $\text{Nb}_2\text{Co}_2\text{GaSb}$, and (d)  p-type $\text{Nb}_2\text{Co}_2\text{GaSb}$.}
 	  \label{fig:zt_1000}
 \end{figure}

The electronic transport properties were combined with lattice thermal conductivity to compute the dimensionless figure of merit ($zT$) as a function of carrier concentration at different temperatures. Figure~\ref{fig:zt_1000} presents the $zT$ values for both electron and hole doping at 1000~K. As shown in the figure, the highest $zT$ for Nb$_{2}$Co$_{2}$InSb is achieved for the SQS2 structure, while for Nb$_{2}$Co$_{2}$GaSb, the OS2 structure exhibits the best performance.
For NbCoSn, the maximum $zT$ value in the n-type case is 0.25 at a carrier concentration of $1.88 \times 10^{21}$~cm$^{\text{-3}}$, whereas in the p-type case, it reaches 0.10 at $4.35 \times 10^{21}$~cm$^{\text{-3}}$.
In Nb$_{2}$Co$_{2}$InSb (n-type), the peak $zT$ values for OS1, OS2, SQS1, and SQS2 are 0.98, 0.99, 1.05, and 1.55 at carrier concentrations of $1.19 \times 10^{21}$, $8.36 \times 10^{20}$, $7.72 \times 10^{20}$, and $9.32 \times 10^{20}$~cm$^{\text{-3}}$, respectively. For the p-type case, the corresponding peak values are 0.79, 1.44, 0.35, and 2.36 at $1.57 \times 10^{21}$, $2.56 \times 10^{20}$, $1.20 \times 10^{21}$, and $6.64 \times 10^{20}$~cm$^{\text{-3}}$.
In the case of Nb$_{2}$Co$_{2}$GaSb (n-type), OS1, OS2, and SQS1 show maximum $zT$ values of 1.85, 2.35, and 2.01 at carrier concentrations of $6.54 \times 10^{20}$, $2.95 \times 10^{20}$, and $4.87 \times 10^{20}$~cm$^{\text{-3}}$, respectively. For the p-type case, the highest $zT$ values are 1.08, 2.09, and 0.54 for OS1, OS2, and SQS1 at carrier concentrations of $7.23 \times 10^{20}$, $4.30 \times 10^{20}$, and $1.21 \times 10^{21}$~cm$^{\text{-3}}$, respectively.

\begin{figure}[h!]
 	\centering
 	\includegraphics[width=\columnwidth]{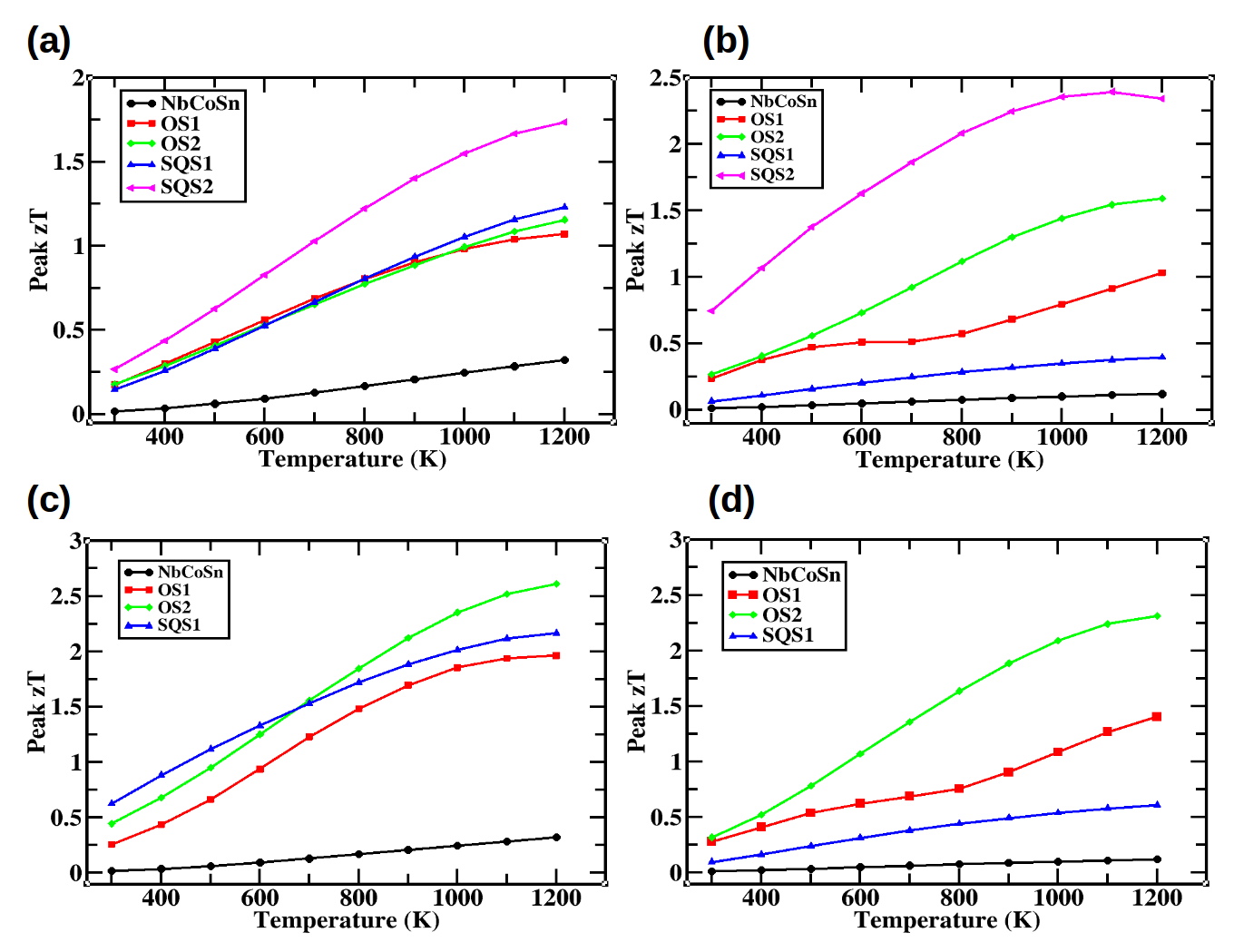} 
 	\caption{ Peak value of figure of merit as a function of temperature of (a) n-type  $\text{Nb}_2\text{Co}_2\text{InSb}$, (b)  p-type $\text{Nb}_2\text{Co}_2\text{InSb}$ , (c) n-type $\text{Nb}_2\text{Co}_2\text{GaSb}$, and (d)  p-type $\text{Nb}_2\text{Co}_2\text{GaSb}$.}
 	  \label{fig:peak_zt_ingasb}
 \end{figure}

Figure~\ref{fig:peak_zt_ingasb} presents the peak $zT$ values as a function of temperature. Among the different structures of \( \text{Nb}_2\text{Co}_2\text{InSb} \), SQS2 exhibits the highest thermoelectric performance. In contrast, for \( \text{Nb}_2\text{Co}_2\text{GaSb} \), the OS2 configuration outperforms the other structures.
As shown in Figure~\ref{fig:peak_zt_ingasb}, at 1200~K, the peak $zT$ values for SQS2 of \( \text{Nb}_2\text{Co}_2\text{InSb} \) are 1.73 for n-type and 2.34 for p-type doping. For OS2 of \( \text{Nb}_2\text{Co}_2\text{GaSb} \), the corresponding values at the same temperature are 2.61 (electrons) and 2.31 (holes). These are notably higher than those for \( \text{NbCoSn} \), which reaches a maximum $zT$ of only 0.32 for n-type and 0.12 for p-type carriers at 1200~K.

 \subsection{Conclusion}

 Apart from SQS2 of \( \text{Nb}_2\text{Co}_2\text{InSb} \), all the structures of both the systems are dynamically stable.
The lattice thermal conductivity of these structures is approximately one-fifth that of the ternary half-Heusler compound \( \text{NbCoSn} \). This reduction is attributed to mass and strain field fluctuation scattering, which has a more pronounced impact on the ordered structures compared to the disordered ones.

A high figure of merit ($zT$) was observed for these materials, with values of 1.73 (2.34) for electrons (holes) in the SQS2 structure of \( \text{Nb}_2\text{Co}_2\text{InSb} \), and 2.61 (2.31) for electrons (holes) in the OS2 structure of \( \text{Nb}_2\text{Co}_2\text{GaSb} \) at 1200~K. These $zT$ values are significantly higher than the corresponding values for \( \text{NbCoSn} \), which are 0.32 (0.12) for electrons (holes) at the same temperature.

Due to the high $zT$ values for both electrons and holes, these materials are suitable for use as both \(n\)-type and \(p\)-type legs in thermoelectric devices.

\section*{Acknowledgements}
I would like to sincerely thank Dr.~Prasenjit Ghosh for his constant guidance and support, especially in helping me organize the results and shape the manuscript, and for providing access to the necessary computational resources. I am also grateful to the National Supercomputing Mission (NSM) for access to the \textquotedblleft Param Brahma\textquotedblright\ supercomputer at IISER Pune. I would like to thank Dr.~Gautam Sharma from Khalifa University for many helpful and insightful discussions throughout this work. I also acknowledge IISER Pune for providing financial support in the form of a fellowship.

 \clearpage
 
  \bibliography{references}
\end{document}